\documentclass[aps,prl,twocolumn,superscriptaddress]{revtex4}
\usepackage{amsmath,amssymb}
\usepackage{graphicx}
\usepackage[usenames]{color}
\usepackage[bookmarks=true,colorlinks,linkcolor=red,urlcolor=blue,citecolor=blue]{hyperref}

\input epsf
\usepackage{epstopdf}

\definecolor{darkgreen}{rgb}{0,0.4,0}
\definecolor{darkred}{rgb}{0.4,0,0}
\definecolor{darkblue}{rgb}{0,0,0.4}

\def\be{\begin{equation}}
\def\ee{\end{equation}}
\newcommand{\bea}{\begin{eqnarray}}
\newcommand{\eea}{\end{eqnarray}}

\input epsf

\newlength{\extraspace}
\newlength{\extraspaces}
\def\ket#1{|#1\rangle}

\def\II{\relax{I\kern-.10em I}}

\def\IZ{\relax{\rm Z\kern-.34em Z}}
\def\IB{\relax{\rm I\kern-.18em B}}
\def\IC{{\relax\hbox{$\inbar\kern-.3em{\rm C}$}}}
\def\ID{\relax{\rm I\kern-.18em D}}
\def\IE{\relax{\rm I\kern-.18em E}}
\def\IF{\relax{\rm I\kern-.18em F}}
\def\IG{\relax\hbox{$\inbar\kern-.3em{\rm G}$}}
\def\IGa{\relax\hbox{${\rm I}\kern-.18em\Gamma$}}
\def\IH{\relax{\rm I\kern-.18em H}}
\def\II{\relax{\rm I\kern-.18em I}}
\def\IK{\relax{\rm I\kern-.18em K}}
\def\IP{\relax{\rm I\kern-.18em P}}
\def\inbar{\,\vrule height1.5ex width.4pt depth0pt}

\def\IR{\relax{\rm I\kern-.18em R}}

\def\lp10{\ell_p^{10}}
\def\lp11{\ell_p^{11}}
\def\R11{R_{11}}

\def\frac#1#2{{#1 \over #2}}

\newdimen\tableauside\tableauside=1.0ex
\newdimen\tableaurule\tableaurule=0.4pt
\newdimen\tableaustep
\def\phantomhrule#1{\hbox{\vbox to0pt{\hrule height\tableaurule width#1\vss}}}
\def\phantomvrule#1{\vbox{\hbox to0pt{\vrule width\tableaurule height#1\hss}}}
\def\sqr{\vbox{%
  \phantomhrule\tableaustep
  \hbox{\phantomvrule\tableaustep\kern\tableaustep\phantomvrule\tableaustep}%
  \hbox{\vbox{\phantomhrule\tableauside}\kern-\tableaurule}}}
\def\squares#1{\hbox{\count0=#1\noindent\loop\sqr
  \advance\count0 by-1 \ifnum\count0>0\repeat}}
\def\tableau#1{\vcenter{\offinterlineskip
  \tableaustep=\tableauside\advance\tableaustep by-\tableaurule
  \kern\normallineskip\hbox
    {\kern\normallineskip\vbox
      {\gettableau#1 0 }%
     \kern\normallineskip\kern\tableaurule}%
  \kern\normallineskip\kern\tableaurule}}
\def\gettableau#1 {\ifnum#1=0\let\next=\null\else
  \squares{#1}\let\next=\gettableau\fi\next}

 \def\eqnn#1{\xdef #1{(\secsym\the\meqno)}\writedef{#1\leftbracket#1}%
 \global\advance\meqno by1\wrlabeL#1}
 \def\eqna#1{\xdef #1##1{\hbox{$(\secsym\the\meqno##1)$}}
 \writedef{#1\numbersign1\leftbracket#1{\numbersign1}}%
 \global\advance\meqno by1\wrlabeL{#1$\{\}$}}
 \def\eqn#1#2{\xdef #1{(\secsym\the\meqno)}\writedef{#1\leftbracket#1}%
 \global\advance\meqno by1$$#2\eqno#1\eqlabeL#1$$}

\global\newcount\itemno \global\itemno=0

\def\itemaut#1{\global\advance\itemno by1\noindent\item{\the\itemno.}#1}

\def\({\left(}
\def\){\right)}

\def\ie{{\it i.e.}}

\usepackage[yyyymmdd,hhmmss]{datetime}
\newif{\ifeq}           % defines a new condition @eq tested by the conditional \ifeq
\eqtrue                 % if uncommented, declares @eq to be true
\begin{document}

\title{Polynomial Monogamy Relations for Entanglement Negativity}
\author{Grant W. Allen}
\affiliation{Department of Physics, University of California at San Diego, La Jolla, CA 92093-0354} 
\author{David A. Meyer}
\affiliation{Department of Mathematics, University of California at San Diego, La Jolla, CA 92093-0112}

\begin{abstract}
The notion of non-classical correlations is a powerful contrivance for explaining phenomena exhibited in quantum systems. It is well known, however, that quantum systems are not free to explore arbitrary correlations---the church of the smaller Hilbert space only accepts monogamous congregants. We demonstrate how to characterize the limits of what is quantum mechanically possible with a computable measure, entanglement negativity. We show that negativity only saturates the standard linear monogamy inequality in trivial cases implied by its monotonicity under LOCC, and derive a necessary and sufficient inequality which, for the first time, is a non-linear higher degree polynomial. For very large quantum systems, we prove that the negativity can be distributed at least linearly for the tightest constraint and conjecture that it is at most linear.

\end{abstract}
%\pacs{Keywords: }
\maketitle

%\tableofcontents

The prototypical quantum correlation, entanglement, accounts for many of the effects seen in quantum theory, and there is a significant effort to enslave it as a resource to be manipulated and exploited by quantum engineers~\cite{HHHH}. The most striking property of entanglement is its distributability, or lack there of, exemplified by its compliance with, for example, monogamy laws~\cite{OV06} and area laws~\cite{ECP}, laws which constrain the shareability of correlations. The former law states that the closer two parties are to being maximally entangled, the more the pair become separated and hidden from all other parties. The latter law requires that for certain types of many-particle ground states split in twain, the entanglement between the parts scales according to the size of the separation boundary. Both laws set the stage for entanglement to play a central role in physics, and an active research community continues to strengthen this role~\cite{Wen}.

The secludedness of maximally entangled states has specifically impacted quantum key distribution~\cite{P10}, ground state frustration~\cite{FGA07}, and even black holes~\cite{VV12}. These perspectives on monogamy have traditionally been understood in terms of multi-qubit networks. In the case of tripartite systems, the monogamy law can be written in terms of concurrence~\cite{CKW00}, as the following,
\begin{equation}\label{mono}
\mathcal{C}_{A|BC}^2\geq \mathcal{C}_{A|B}^2+\mathcal{C}_{A|C}^2,
\end{equation}
where $\mathcal{C}$ is the concurrence, subscripts label the parties, and the vertical bar denotes the bipartite split across which it is computed.  In this way, if $\mathcal{C}_{A|BC}^2 \approx \mathcal{C}_{A|B}^2 \approx 1 $, \ie, $A$ and $B$ are maximally entangled, then $\mathcal{C}_{A|C}^2 \approx 0$, hence the personification of entanglement. 

The elegance and simplicity of the monogamy inequality has made it the paragon for entanglement shareability.  It has since been shown that other entanglement measures such as entanglement of formation~\cite{OCF14}, squashed entanglement~\cite{KGS}, and entanglement negativity~\cite{OH}, all satisfy the same monogamy relation. This raises two issues: the monogamy inequality may just be a first order approximation to the shareability of correlational resources, so to what extent can we quantify the exact amount possible to share? Second, is the limited shareability a consequence of limited systems, \ie, qubits?

The first issue has recently been under pressure through a few angles. For example, it is well-known that the ring of polynomial invariants is finitely generated, and an explicit algebraic dependency, \ie, a polynomial, was discovered, relating all linear entropies, and hence coined, an exact monogamy \textit{equality}~\cite{ESmono}. Other considerations further refute the sanctity in the linear monogamy inequality \eqref{mono} claiming that it does not suffice to capture all entanglement trade-off constraints, constraints which rather may be unique to each entanglement \textit{measure}~\cite{WAP16}.

\begin{figure}[ht]
	\includegraphics[width=.3 \textwidth]{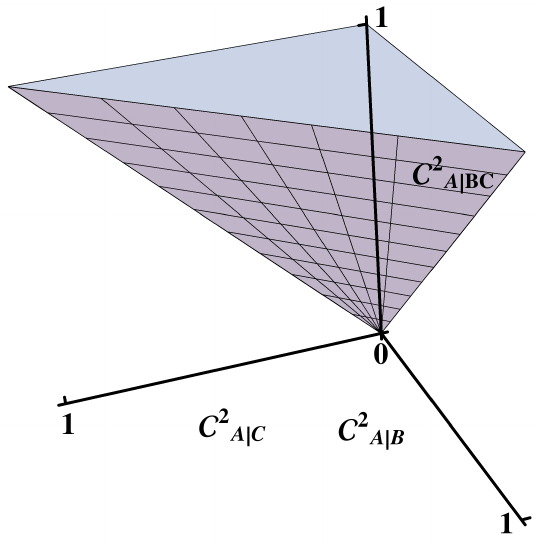}
	\caption{Achievable qubit concurrence}
	\label{conimage}
\end{figure}

In any application of monogamy, it is then important to note that {\textit{some}} law being satisfied does not necessarily imply its physical achievability. Concurrence is an exception to this rule: consider the achievable set of triples, $(\mathcal{C}_{A|C}^2, \mathcal{C}_{A|B}^2,\mathcal{C}_{A|BC}^2)$, computed for all pure states of 3 qubits.  The equality in \eqref{mono} then defines a plane in the ambient space, $[0,1]^3$.  The original work~\cite{CKW00} found saturation with the $W$ class of states, $\ket{W} = a \ket{001} + b\ket{010} + c \ket{100}$, which do, in fact, map onto the entire plane in $[0,1]^3$.  Less well-known is the fact that the entire region above the plane can be achieved; see Fig.~\ref{conimage}.

Entanglement negativity is not an exception to the above achievability rule. One aim of this Letter is to make the fine print of its monogamy law legible; in other words, to derive the boundary of achievable correlations for negativity. Negativity, in particular, is important for several reasons. It is directly related to PPT states, a peculiar set of entangled states that, among other properties~\cite{AJS14}, cannot be distilled~\cite{VW02}, so that negativity may be measuring ``useful" entanglement.  Further, negativity provides an alternative measure of mixed state entanglement that has the extremely rare property of being computable, indeed for \textit{arbitrary} dimensional quantum system.

Recent advances in quantum photonics have allowed large dimensional subsystems to become common place~\cite{BBFS13}. There has even been a confirmation of entanglement between $D=100$ dimensional qudits~\cite{Z14}. The need for a better understanding of correlations in these systems has arrived and negativity is one route to reach it. Negativity further rears its head in the study of gaussian states and continuous variable entanglement~\cite{AI07}. 

Higher dimensional monogamy seemed to have had a wrench thrown into the works when the  3-qutrit Levi-Civita state led to a violation of the standard monogamy inequality \eqref{mono} for the I-concurrence~\cite{O07}. On the other hand, the negativities of the Levi-Civita state evaluates safely within the bounds of the same inequality~\cite{KDS09}. Such a scenerio touches exactly upon the question of Lancien \textit{et al.}, whether entanglement measures be monogamous or faithful~\cite{WAP16}, a trade-off which negativity just might have negotiated.  Therefore, another aim of this Letter is to touch on large quantum systems, for which we give a lower bound to the negativity achievability boundary in all dimensions. Numerical evidence suggests it to be the upper bound as well.

Negativity is based on the failure of the transpose operation to preserve positivity when acting on subsystems~\cite{VW02}.  Transposing a separated system leaves the positivity unaffected, so a state with a non-positive partial transpose must be entangled. Negativity is defined as twice the sum of the negative eigenvalues of the partially transposed state:
\begin{equation}\label{Ndef}
\mathcal{N}_{A|B} = \mathcal{N}(\rho_{AB}) = 2 \sum_{n} \lambda^{(-)}_n(\rho^{T_A}_{AB}), 
\end{equation}
where $\rho^{T_A}_{AB}$ is the density matrix with the $A$ tensor factor transposed (the factor of 2 gives the bell states unit negativity). Although negativity is not itself a convex roof construction, it remarkably has many properties expected of an entanglement measure, \textit{e.g.}, it has an interpretation as a distance to a separable state~\cite{KKS07}.  One can straightforwardly verify that for pure $2\times2$ systems, negativity and concurrence agree, $\mathcal{N} = \mathcal{C}$, and due to the convexity of negativity \cite{VW02}, the monogamy inequality for negativity immediately follows.

In order to find the tighter monogamy inequality, \ie, the boundary of the achievable set $(\mathcal{N}_{A|C}^2, \mathcal{N}_{A|B}^2,\mathcal{N}_{A|BC}^2)$, it will be useful to have a parametrization for the 3-qubit pure states.  Ac{\'\i}n \textit{et al.}\ showed how to ``rotate out'' all local unitaries to achieve a canonical form---a tripartite analog to Schmidt decomposition for pure bipartite states~\cite{AAJT}.  One such form is given as
\begin{equation}\label{acin}
\ket{\Psi} = d \ket{000} + \omega  \ket{100}+ a \ket{101}+b \ket{110} + c  \ket{111}
\end{equation}
with real $a,b,c,d\geq0$, $\omega \in \mathbb{C}$, and the usual normalization. The boundary will be computed from ``below'' by maximizing the negativities in this parametrization.

Qubit negativity necessarily satisfies $0\leq \mathcal{N} \leq 1$. We find the one-two party split negativity straight forwardly,
\begin{equation*}
\mathcal{N}_{A|BC}^2=4(a^2 + b^2 +c^2) d^2=4(1-d^2-|\omega|^2) d^2,
\end{equation*}
where the last equality follows from the normalization constraint. By maximizing with respect to $d$, we find the inequality, $0\leq \mathcal{N}_{A|BC}^2  \leq (1-|\omega|^2)^2$.  Thus when $\mathcal{N}_{A|BC}^2$ is maximal, the parameter $\omega$ will vanish. For $\mathcal{N}_{A|BC}^2$ not maximal, it will be useful to consider for what values of $\omega$ are $\mathcal{N}_{A|C}^2$ and $\mathcal{N}_{A|B}^2$  both maximized. 

To calculate $\mathcal{N}_{A|C}^2$, we need the following fact:  a partial transpose cannot produce more than $(D-1)^2$ negative eigenvalues for two entangled $D$-dimensional systems~\cite{R13}.  Accordingly, for two-qubit states, $\rho^{T_A}_{AC}$ has no more than one negative eigenvalue.  The negativity is twice the negative eigenvalue, and thus satisfies the quartic polynomial equation, 
\begin{align}
0 &= \text{det} (2\rho^{T_A}_{AC}+\mathcal{N}_{A|C}I_4) \nonumber\\
  &= -16 a^4 d^4 - 16 a^2 c^2 d^4 - 8 a^4 d^2 x + 8 a^2 b^2 d^2 x \nonumber\\
  &\phantom{=}\; -8 a^2 c^2 d^2 x -8 a^2 d^4 x - 8 a^2 d^2 |\omega|^2 x + 4 a^2 b^2 x^2 \nonumber\\
  &\phantom{=}\; + 4 b^2 d^2 x^2 +  4 c^2 d^2 x^2 +4 c^2 |\omega|^2 x^2 + 2 a^2 x^3 \nonumber\\
  &\phantom{=}\; + 2 b^2 x^3 + 2 c^2 x^3 + 2 d^2 x^3 +2 |\omega|^2 x^3 + x^4 \nonumber\\
  &\phantom{=}\; - (16 d^2 x +  8 x^2) a b c |\omega| \cos(\text{arg}(\omega)), \label{quartic}
\end{align}
where $I_4$ is the $4\times4$ identity matrix and $x = \mathcal{N}_{A|C}$.  Implicitly differentiating this quartic with respect to arg$(\omega)$ and setting $\partial x/ \partial(\text{arg}(\omega))=0$ gives
\begin{equation*}
8 a b c |\omega| x (2 d^2 + x) \sin(\text{arg}(\omega)) =0,
\end{equation*}
so henceforth we restrict $\omega \in \mathbb{R^{+}}$ and drop the absolute value; the potential extra minus sign from $\omega \in \mathbb{R^{-}}$ does not affect the end result.  Once again, differentiating the quartic in \eqref{quartic} with respect to $\omega$ and setting $\partial x/\partial\omega=0$ gives
\begin{equation}\label{constraint1}
x (4 a b c d^2 + 4 a^2 d^2 \omega + 2 a b c x - 2 c^2 \omega x - \omega x^2) =0.
\end{equation}
The gives one constraint, and along with normalization, leaves only three parameters. Since we are after a boundary surface in $[0,1]^3$, we will eliminate another variable: Using \eqref{quartic} to maximize $x=\mathcal{N}_{A|C}$ with respect to $c$ gives:
\begin{multline}\label{constraint2}
8 a^2 c d^4 + 4 a^2 c d^2 x + 4 a b d^2 \omega x - 2 c d^2 x^2 \\+  2 a b \omega x^2 - 2 c \omega^2 x^2 - c x^3=0.
\end{multline}
Now we employ a powerful technique from computational algebraic geometry to perform algebraic elimination.  Finding the minimal generating set, the \textsl{Gr\"obner basis}, for the ideal generated by these polynomial constraints \cite{hibi}, will give polynomials with the proper variables eliminated.  The Gr\"obner basis for Eqs.~\ref{quartic} (with $\omega\in\mathbb{R}$), \ref{constraint1} and \ref{constraint2} has a single element; setting it to zero gives:
\begin{multline} \label{xsols}
(2 a d - x) (2 a^2 + x) (2 a d + x) (2 d^2 + x) \\
\times (2 a^2 d^2 + a^2 x -  b^2 x)^2 (4 a^2 d^2 - 2 b^2 x - x^2)=0.
\end{multline}
Neglecting negative solutions for $x$ leaves three options: $x = 2 a d$, and $x=-b^2 + \sqrt{b^4 + 4 a^2 d^2}$, the former producing a sub-manifold of the latter with $b=0$. The solution $x =2 a^2 d^2 / (b^2-a^2)$ also produces a sub-manifold: to show this, we now enforce normalization and find one more Gr\"obner basis. Eliminating $x$, $\omega$, $c$, given the following constraints, $x =2 a^2 d^2 / (b^2-a^2)$, Eqs.~\ref{quartic} (with $\omega\in\mathbb{R}$), \ref{constraint1}, and normalization, produces,
\begin{equation}
a^7 b^3 d^4 (a^2 - b^2 - a d) (a^2 - b^2 + a d) (-a^2 + b^2 +  2 a^2 d^2)^2 =0,
\end{equation} 
demonstrating that $x =2 a^2 d^2 / (b^2-a^2) = 2 a d$ is again a sub-manifold of solution $x=-b^2 + \sqrt{b^4 + 4 a^2 d^2}$.  A similar analysis on $\mathcal{N}^2_{A|B}$ leads to the following triples,
\begin{equation} \label{Wnegs1}
\begin{pmatrix} \mathcal{N}^2_{A|C}  \\ \mathcal{N}^2_{A|B} \\ \mathcal{N}^2_{A|BC} \end{pmatrix} = 
\begin{pmatrix}  \left(b^2 - \sqrt{b^4 + 4 a^2 d^2}\right)^2 \\ \left(a^2 - \sqrt{a^4 + 4 b^2 d^2}\right)^2 \\ 4 (a^2 + b^2) d^2  \end{pmatrix} .
\end{equation}
These triples come from precisely the condition that $\omega = c = 0$, leaving states in \eqref{acin} that are locally equivalent to the $W$ class, the same class that maximize concurrence.  The three components of \eqref{Wnegs1} parametrically define the boundary of the achievable set. Together with the normalization constraint, we can eliminate the state coefficients, turning the parametric surface into an implicit surface.  Computing the Gr\"obner basis of the parametric polynomials, we again find a single element, which set to $0$ gives the surface implicitly:
\begin{align} 
& z^6  - 2 z^4 (x^2 - x y + y^2) \nonumber\\
&\,+ z^2 \Bigl(x^4 +y^4 - 2x y\bigl(x(x-1)+y(y-1)- 3x y/2 + 2\bigr)\Bigr) \nonumber\\
&\,+ x y (2 y^2 + x y^2 +  x^2 y + 2 x^2 )(x + y + 2) = 0,
\label{bound}
\end{align}
where we identify $(x^2,y^2,z^2)\equiv(\mathcal{N}_{A|C}^2 ,\mathcal{N}_{A|B}^2, \mathcal{N}_{A|BC}^2)$. 

\begin{figure}[ht]
	\includegraphics[width=.35 \textwidth]{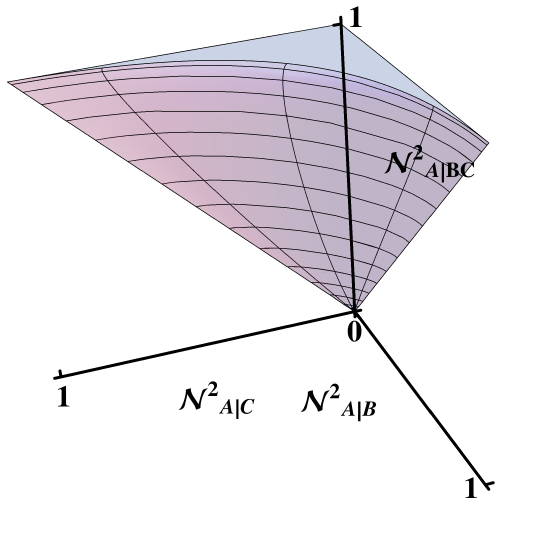}
	\caption{Achievable qubit negativity}
	\label{negimage}
\end{figure}

We now show that \eqref{bound} is the only non-trivial boundary of the achievable set. Adding back in the parameter $c$ will fill in the rest of the set.  The partially transposed reduced states are then full rank so that the determinants are negative as long as there is entanglement,
\begin{align*}
\det \rho_{A|C}^{T_A} &= - a^2 d^4 (a^2 + c^2), \\
\det \rho_{A|B}^{T_A} &= - b^2 d^4 (b^2 + c^2).
\end{align*}
Since $\mathcal{N}_{A|BC}^2=4(1-d^2) d^2$, fixing $z^2$ will fix $d$.  Thus, on a constant $z^2$ plane, starting at the boundary ($c=0$), which is a curve intersecting the $x^2z^2-$ and $y^2z^2-$coordinate planes at $(z^2,0,z^2)$ and $(0,z^2,z^2)$, respectively, increasing $c$ to $\sqrt{1-d^2}$ will smoothly collapse the curve into the point $(0,0,z^2)$ as $a$ and $b$, and hence the determinants and the negativities, vanish. During this collapse, the curve continues to intersect the coordinate planes and traces out the achievable $z^2$-plane set. As $z^2$ is arbitrary, all points between the boundary and the $z^2$-axis can be achieved.  See Fig.~\ref{negimage} for the achievable negativity set.

We would like to describe the achievable negativity set with a single inequality, as is done with achievable concurrence set. It is not enough to modify \eqref{bound} to be non-positive, simply due to the existence of multiple solutions $z=z(x,y)$, one of which intersects the achievable set's interior. We can factor out the unwanted solution with the following coordinate transformation: $z \rightarrow 2\sqrt{\lambda (1-\lambda)}$. Plugging in to \eqref{bound}, the polynomial factors,
\begin{align}
&\big(8 \lambda^3 -  16 \lambda^2 \\
& +2 (x^2+y^2-x y +4) \lambda  - (2x^2 +2y^2 +x^2 y + x y^2) \big) \nonumber \\
& \times (8 \lambda^3 -8\lambda^2+ 2 (x^2+y^2-xy)\lambda + (x^2 y + x y^2 + 2 x y) ).\nonumber
\end{align}
We take only the 2nd factor, and transform coordinates back, $\lambda \rightarrow  \frac{1}{2}(1+\sqrt{1-z^2})$, which leaves us with the necessary and sufficient monogamy inequality for achievable negativities,

\begin{align}
(z^2 - x^2 -y^2 +x y)(1+ \sqrt{1-z^2} ) -x y(2+x+y)\geq 0.
\end{align}

The states whose negativities fill up the region shown in Fig.~\ref{negimage} can be understood to have a special form, which is useful for generalizing this result to higher dimensional tensor factors. From \eqref{acin} with $\omega=0$,
\begin{align*}
\ket{\Psi} &= d \ket{000} +  a \ket{101}+b \ket{110} + c  \ket{111} \\ 
           &= a \ket{\Phi}_{AC}\ket{0}_B + b \ket{\Phi}_{AB}\ket{0}_C + c \ket{\text{GHZ}}_{ABC} \\
           &\quad + (d-a-b-c)\ket{000},
\end{align*}
where $\ket{\Phi} = \ket{00}+\ket{11}$ and $\ket{\text{GHZ}}=\ket{000}+\ket{111}$.  These states then generalize to $D$-dimensional qudits straightforwardly \textit{via\/} $\ket{\Phi} \rightarrow \sum_j \ket{jj}$ and  $\ket{\text{GHZ}} \rightarrow \sum_j \ket{jjj}$,
\begin{align}\label{dit}
\ket{\Psi} = d \ket{000} + \sum_{j=1}^{D-1}  a\ket{j0j} + b\ket{jj0}+c \ket{jjj}.
\end{align}
The partial transpose of the reduced density operator for \eqref{dit} block diagonalizes to:
\begin{equation*}
\rho_{A|C}^{T_A} = \begin{pmatrix}d^2\end{pmatrix}\bigoplus_{j=1}^{\frac{(D-1)(D-2)}{2}}\begin{pmatrix} 0 & a^2 \\ a^2 & 0\end{pmatrix} \bigoplus_{j=1}^{D-1}  \begin{pmatrix} 0 & a d & 0 \\ a d  & b^2 & b c \\ 0 & b c & a^2 + c^2 \end{pmatrix}.
\end{equation*}
$\rho_{A|B}^{T_A}$ has the same form with $a$ and $b$ interchanged.  The similarity with the $D=2$ case, particularly the $3\times 3$ matrix factor, tells us that setting $c=0$ maximizes the pairwise negativities for this family. The third negativity can be computed to be:
\begin{multline*}
\mathcal{N}_{A|BC} = 2 (D-1) d \sqrt{a^2+b^2+c^2} \\+ (D-1)(D-2) (a^2+b^2+c^2),
\end{multline*}
so that again, fixing $d$ will fix $\mathcal{N}_{A|BC}$. The same argument about the determinants applies again, so 
\begin{align*}
\det \rho_{A|C}^{T_A} &= (-1)^{\lfloor \frac{D}{2} \rfloor} d^2 \bigl((a^2 + c^2)d^2\bigr)^{(D-1)}a^{2(D-1)(D-2)} \\
\det \rho_{A|B}^{T_A} &=  (-1)^{\lfloor \frac{D}{2} \rfloor} d^2 \bigl((b^2 + c^2)d^2\bigr)^{(D-1)}b^{2(D-1)(D-2)},
\end{align*}
on a constant $\mathcal{N}_{A|BC}$-plane.  And as before, increasing $c$ to $\sqrt{(1-d^2)/(D-1)}$ will send $a$ and $b$ to 0 so the determinants, and thus the pairwise negativities, vanish.  Since this is a natural extension of the achievable region we found for $D = 2$, we conjecture that it is the entire achievable set of negativities, $(\mathcal{N}_{A|C}^2 ,\mathcal{N}_{A|B}^2, \mathcal{N}_{A|BC}^2 )$, for $D>2$ as well.

%%%%%%%%%%%%%%%%%%%%%%%%%%%%%%%%%%%%%%%%%%%%%%%%
For the boundary states ($c=0$) the parameters can be eliminated for the Gr\"obner basis to get the conjectured implicit bound; however already in the $D=3$ case, the polynomial is rather complicated, containing 143 terms.  It is worth mentioning that na{\"\i}vely testing the conjecture numerically is nearly hopeless, since the negativities of random states are highly non-uniform throughout the achievable set \cite{Datta}.  Nevertheless, testing for perturbations of our boundary has led to no counter-examples. 

An alternative way to derive our boundary states is the following.  Consider the class of maximally entangled states between $A$ and $B$, with an ancillary qudit, $C$,
\begin{equation}\label{maxzero}
\ket{\Psi} = \Bigl( d \ket{00} + \sum_{j=1}^{D-1} b\ket{jj} \Bigr) \ket{0}.
\end{equation}
These states give the entire line, $\mathcal{N}_{A|B} = \mathcal{N}_{A|BC}$ with $\mathcal{N}_{A|C} =0$.  Furthermore, given the two-qudit swap Hamiltonian, 
\begin{equation*}
H_{\text{SWAP}} = \bigoplus_{j\le k} \sigma_x^{(jk)},
\end{equation*}
where $\sigma_x^{(j k)}$ is $\sigma_x$ acting in the $\{j,k\}$-subspace (unless $j=k$ in which case it is just 1 acting on the $\{j\}$-subspace), let it act on the $B$ and $C$ qudits in \eqref{maxzero}:
\begin{align}\label{swap}
e^{i \theta H_{\text{SWAP}}}\ket{\Psi} &= d e^{i \theta} \ket{000} \nonumber\\
&\quad + b \sum_{j=1}^{D-1}\cos{(\theta)} \ket{jj0} +   i \sin{(\theta)}  \ket{j0j}.
\end{align}
Then the phases can be cleaned up with local unitaries to match \eqref{dit}, with $c=0$.

\begin{figure}[ht]
	\includegraphics[width=.25 \textwidth]{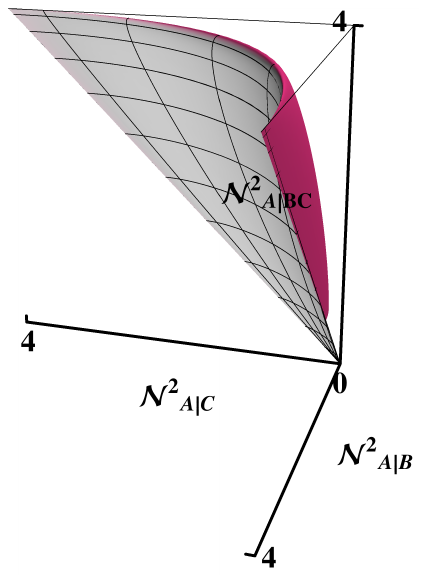}
	\caption{Achieved $D=3$ negativity of states in Eq. \ref{swap}.  Red indicates the (nonbounding) part of the polynomial surface for $d < 1/\sqrt{D}$.}
	\label{trits}
\end{figure}

Note that the boundary states \eqref{swap} produce a surface that extends and folds back into the achievable set depending on the value of $d$, \ie, only maximizing the pairwise negativities when $d>1/\sqrt{D}$, as seen from parametrically plotting the resulting polynomial surface in Fig.~\ref{trits}.  Recall that negativity for $D\times D$ systems has bounds $0\leq \mathcal{N} \leq D-1$ \cite{ES14}.

The condition on $d$ is related to conditions on the marginal eigenvalues:  Higuchi found a necessary condition on the univariate marginal eigenvalues for pure $N$-qudit systems~\cite{H03}; for three qudits it is,
\begin{equation}
\sum_{n=1}^{D-1}\lambda^{(A)}_{n} \leq \sum_{n=1}^{D-1}\lambda^{(B)}_{n} +\sum_{n=1}^{D-1}\lambda^{(C)}_{n},
\end{equation}
including permutations of the parties, where $\lambda^{(P)}_n\leq\lambda^{(P)}_{n+1}$, $n\in\{1,\ldots,D-1\}$ are the ascending eigenvalues of party $P$'s state.  The marginal eigenvalues of \eqref{dit} with $c=0$ are
\begin{align*}
\lambda^{(A)} &= \{ (a^2 + b^2)_{(D-1)}, d^2\} \\
\lambda^{(B)} &= \{ (b^2)_{(D-1)} , (D-1) a^2 + d^2\} \\
\lambda^{(C)} &= \{ (a^2)_{(D-1)} , (D-1) b^2 + d^2\},
\end{align*}
where the subscripts denote the degeneracy.  When $d > 1/\sqrt{D}$, the remaining, smaller, eigenvalues saturate the marginal inequality:
\begin{equation*}
(D-1)(a^2 + b^2) \leq (D-1) a^2 + (D-1) b^2.
\end{equation*}
The expressions for the negativities are similar, to leading order in $D$: 
\begin{equation} \label{Wnegs}
\begin{pmatrix} \mathcal{N}_{A|C}  \\ \mathcal{N}_{A|B} \\ \mathcal{N}_{A|BC} \end{pmatrix} = 
D^2 \begin{pmatrix}  a^2\\ b^2 \\ a^2+b^2 \end{pmatrix} + O(D).
\end{equation}
If our conjecture about achievable negativities is true, then in the limit of large dimensions, the monogamy inequality simplifies to
\begin{equation*}
\mathcal{N}_{A|BC} \geq \mathcal{N}_{A|B}+\mathcal{N}_{A|C},
\end{equation*}
up to terms of $O(1/D)$.  

In summary, we've seen that although negativity is not itself a polynomial invariant in the state coefficients, the qubit negativity satisfies a polynomial equation.  Simple maximization procedures combined with application of Gr\"obner basis computations enabled us to derive an explicit expression for the boundary of the achievable set, a polynomial surface.  Generalizing the qubit boundary states motivated a conjecture for the boundary of the achievable set for arbitrary dimensional qudits, a conjecture supported by numerical experimentation.  We expect our approach to qubits to be relevant in proving the conjecture for qutrits.  For arbitrary dimensions we suspect an intimate connection with marginal eigenvalue constraints since our boundary states saturate the eigenvalue boundaries.

\bibliographystyle{h-physrev}
\bibliography{refs}

\end{document}